\documentclass[aps,prd,preprint,floatfix,showpacs,showkeys,letterpaper]{revtex4-1}
\pdfoutput=1
\usepackage[pdftex]{color}
\usepackage{latexsym,amssymb,amsmath,gensymb}
\usepackage{graphicx,dcolumn,bm}
\usepackage{slashed,kec,bigints}
\usepackage{mathtools}

\DeclareMathOperator{\arcsinh}{arcsinh}

\begin{document}
\title[The radiation, matter, and dark-energy eras]{The eras of radiation, matter, and dark energy: new information from the Planck Collaboration}
\author{Kevin Cahill}
\email{cahill@unm.edu}
\affiliation{Department of Physics \& Astronomy,
University of New Mexico\\Albuquerque, New Mexico, USA}
\affiliation{School of Computational Sciences\\
Korea Institute for Advanced Study, Seoul, Korea}
\date{\today}
\begin{abstract}
Data released by the Planck Collaboration in 2015 imply new dates for the era of radiation, the era of matter, and the era of dark energy.  The era of radiation ended, and the era of matter began, when the density of radiation dropped below that of matter.  This happened \( 50,953 \pm 2236 \) years after the time of infinite redshift when the ratio \( a(t)/a_0 \) of scale factors was \( \left( 2.9332 \pm 0.0711 \right) \by 10^{-4} \)\@.  The era of matter ended, and the era of dark energy began, when the density of matter dropped below that of dark energy (assumed constant)\@.  This happened \( \left( 10.1928 \pm 0.0375 \right) \) Gyr after the time of infinite redshift when the scale-factor ratio was \( 0.7646 \pm 0.0168 \)\@.  The era of dark energy started 3.606 billion years ago.
\par
In this pedagogical paper, five figures trace the evolution of the densities of radiation and matter, the scale factor, the redshift, and the luminosity distance  through the eras of radiation, matter, and dark energy.
\end{abstract}

\maketitle

\section{The Planck Data
\label{The Planck Data}}

The Planck Collaboration
and the European Space Agency are
publishing their remarkable data in many articles
on the arXiv and in various journals.
Their paper~\cite{Ade:2015xua} 
on the cosmological parameters
(mainly column 7 of table 4)
contains the data summarized
in the table below.
In particular,
the present value of the Hubble constant is 
\(H_0 = 67.74 \pm 0.46 \) km/(s\,Mpc)\@.
Equivalently, \( H_0 \) is the frequency
\(  H_0 = ( 2.1953 \pm 0.015 ) 
\by 10^{-18} \) s\({}^{-1}\)
\( = (0.06928 \pm 0.00047 ) \) Gyr\({}^{-1}\), 
and the Hubble time is
\( 1/H_0 
= \left( 4.5552 \pm 0.031 \right) \by 10^{17} \) s
\( = 14.435 \pm 0.098 \) billion (tropical) years.   
The present critical energy density is
\( \rho_{c 0} = 3 H_0^2/8\pi G = \left( 8.6197 \pm 0.1171 \right) \by 10^{-27} \) kg\,m\({}^{-3}\),
and the fraction 
that is due to dark energy is
\( \Omega_{\Lambda 0} = 0.6911 \pm 0.0062 \)\@.
The fraction due to matter is
\( \Omega_{m 0} = 0.3089 \pm 0.0062 \)\@.
The  fraction due to ordinary matter
is \( \Omega_b = 0.0486 \pm .0007 \)
or 15.7\% of \(\Omega_m\)\@.
The present ratio of the total energy density to the 
critical energy density is 
\( \Omega_0 = 1.0000 \pm 0.0088 \)\@.
The age of the known universe
is \( t_0 = 13.799 \pm 0.021\) Gyr\@.

\begin{table}[htp]
\begin{center}
\begin{tabular}{|c|c|c|c|}
\hline\hline
\( H_0 \) & \( \rho_{c 0} = \frac{3 H_0^2}{8\pi G} \) & \( t_0 \)  
& \( \Omega_0 \) \\
\( 67.74 \pm 0.46 \) km/(s\,Mpc)  &
\( \left( 8.6197 \pm 0.1171 \right) \by 10^{-27}  \) kg/m\({}^3\)
& \(  13.799 \pm 0.021\) Gyr &  \( 1.0000 \pm 0.0088 \)\\
\hline
\( \Omega_{\Lambda 0} \) & \( \Omega_{m 0} \)
& \( \Omega_{b 0} \) 
& \( \Omega_k  = - \frac{k c^2}{a^2_0 H^2_0} \) \\
\( 0.6911 \pm 0.0062 \) & \( 0.3089 \pm 0.0062 \) 
& \( 0.0486 \pm .0007 \) &  \( 0.0008 \pm 0.004 \) \\
\hline\hline
\end{tabular}
\end{center}
\caption{Cosmological parameters and their standard deviations
\( \sigma \) (68\% confidence level)\@.}
\label {table}
\end{table}

\par
In section~\ref{The density of radiation},
I use the COBE/FIRAS 
measurement~\cite{2009ApJ...707..916F} 
of the temperature of the
cosmic microwave background (CMB) radiation
\( T_{0 \gamma} = 2.7255 \pm 0.0006 \) K 
to compute the present mass density \( \rho_{r 0} \) 
of radiation as \( \rho_{r 0} = \left(
7.8099 \pm 0.0069 \right) \by 10^{-31} \)
kg/m\({}^3\)\@.
In 
section~\ref{How the scale factor and redshift vary with time},
I derive an integral formula for the time
\( t(a/a_0) \) as a function of the ratio \( a(t)/a_0 \)
of the scale factor \( a(t) \) to its present value
\( a_0 \)\@.
The era of radiation ended when the energy
density of matter was the same as that
of radiation.
In section~\ref{The end of the era of radiation},
I estimate that the ratio of scale factors was
\( a(t)/a_0 = 
\left( 2.9332 \pm 0.0711 \right) \by 10^{-4} \)
at the end of the era of radiation, and that
the era of radiation ended \( 50,953 \pm 2236 \) 
years after the time \( t(0) \) of zero scale factor
or infinite redshift \( z = (a_0/a(t)) -1 \)\@.
The era of matter ended when the energy density
of matter dropped to that of dark energy.
In section~\ref{The end of the era of matter},
I estimate that the era of matter ended 
when the scale-factor ratio was
\( a(t)/a_0 = 0.76458 \pm .01681 \), and that
the era of matter ended at
\( t = 10.1928 \pm 0.0375 \) Gyr 
or 3.606 billion years ago.
The luminosity distance \( d_L \) is the distance
that gives the apparent luminosity \( L_a \) in terms
of the absolute luminosity \( L \) as 
\( L_a = L/(4 \pi d^2_L) \)\@.
In section~\ref{The luminosity distance},
I use the Planck data to estimate
the luminosity distance \( d_L(z) \) 
both for moderate redshift (\( z < 10 \))
and for \( 0 < z < 10^6 \);
as \( z \to \infty \), 
the luminosity distance \( d_L(z) \)  rises 
as \( 3.2 \, c \, (1+z)/H_0 \)\@.

\section{The density of radiation
\label{The density of radiation}}

We first compute the
fraction \( \Omega_{r 0} \) of
the present mass density that is due to 
radiation.
The COBE/FIRAS 
measurement~\cite{2009ApJ...707..916F}
of the temperature
of the CMB radiation is
\( T_0 = 2.7255 \pm 0.0006\) K\@.
The mass (energy/\(c^2\)) density
of a gas of photons at that temperature 
is~\cite{Cahill2013.4.10, Weinberg2010.2.1}
\begin{equation}
\rho_{\gamma 0}={} \frac{8 \pi^5 
\left(k_B T_0\right)^4}{15 h^3 c^5}
= \left( 4.6451 \pm  0.0041 \right) 
\by 10^{-31} \,\,\mbox{kg m}^{-3} .
\label{rhophotons}
\end{equation}
The three known neutrinos decoupled
before the photons and therefore
have been cooling longer due to 
the expansion of the universe.
Their present temperature~\cite{Weinberg2010.2.1}  
is \( T_{0 \nu} = (4/11)^{1/3} \, T_0 \)\@.
Adding them in, we estimate
the present mass density of
massless and nearly massless particles as
\begin{equation}
\rho_{r 0} = \left[1 + 3\left(\frac{7}{8}\right)
\left(\frac{4}{11}\right)^{4/3} \right]
\,\, \rho_{\gamma 0} = \left(
7.8099 \pm 0.0069 \right) \by 10^{-31} 
\,\,\mbox{kg m}^{-3} .
\label {rhor}
\end{equation}

\section{How the scale factor and redshift vary with time
\label{How the scale factor and redshift vary with time}}

The first-order Friedmann 
equation~\cite{Cahill2013.11.48, Weinberg2010.1.5}
relates the square of the 
instantaneous Hubble rate \( H = \dot a(t) /a \)
to the mass density \( \rho \) and to the
scale factor \( a(t) \)
\begin{equation}
H^2 = \lt( \frac{\dot a}{a} \rt)^2 
= \frac{8 \pi G}{3} \, \rho - \frac{k c^2}{a^2} 
\label {1st order E eq}
\end{equation}
in which the constant \( k = \pm 1 \) or 0
in suitable coordinates.
Dividing the terms of this equation
(\ref {1st order E eq}) by the square of the 
present Hubble rate \( H_0^2 \),
we get
\begin{equation}
\frac{H^2}{H^2_0} = 
\frac{1}{H^2_0} \lt( \frac{\dot a}{a} \rt)^2 
= \frac{1}{H^2_0} \left(
\frac{8 \pi G}{3} \, \rho - \frac{k}{a^2} \right)
= \frac{\rho}{\rho_{c 0}} - \frac{k c^2}{a^2 H^2_0} .
\label {2d F eq}
\end{equation}
\par
The density \( \rho \) is the sum 
\( \rho = \rho_\Lambda + \rho_r + \rho_m \)
of the vacuum density \( \rho_\Lambda \),
the radiation density \( \rho_r \), and
the matter density \( \rho_m \)\@.
The vacuum density \( \rho_\Lambda \)
may be due to a 
cosmological constant \( \Lambda \)
or it may be the energy density (over \(c^2\)) of
the ground state of the quantum theory
that describes the universe.
In any case, I assume that it is a constant 
\begin{equation}
 \rho_\Lambda = \rho_{\Lambda 0} 
\label {rhov}
\end{equation}
that does not vary with time,
as it would in theories of 
quintessence~\cite{Weinberg2010.1.12, *Peebles:2002gy, *Linder:2007wa}\@.
\par
The critical density \( \rho_{c} = 3 H^2/(8 \pi G) \)
is the one that satisfies the
Friedmann equation (\ref {1st order E eq})
for a flat (\( k = 0 \)) universe.
Its present value is
\begin{equation}
\rho_{c 0} ={} \frac{3 H^2_0}{8 \pi G} 
= \left( 8.6197 \pm 0.1171 \right)
\by 10^{-27} \mbox{ kg\,m}{}^{-3} .
\label {present value of the critical density}
\end{equation}
The present ratio \( \Omega_{r 0} \)
of the radiation density (\ref {rhor})
to the critical density 
(\ref {present value of the critical density}) is then
\begin{equation}
 \Omega_{r 0}= \frac{\rho_{r0}}{\rho_{c 0}}
 = \left( 9.0606 \pm 0.1233 \right)
\by 10^{-5} .
 \label{Omegar}
\end{equation}
Because wavelengths vary 
with the scale factor \( a(t) \),
the radiation density varies as
\begin{equation}
\rho_r(t) = \rho_{r 0} \, \frac{a^4_0}{a^4(t)} .
\label {rhor at time t}
\end{equation}
\par
Baryons have partial lifetimes that exceed
\( 5.8 \by 10^{29} \) years for 
\( n \to \) invisible~\cite{Araki:2005jt},
and \( 2.1 \by 10^{29} \) years
for \( p \to \) invisible~\cite{PhysRevLett.92.102004},
both of which are much longer than 
the age \( 13.8 \) Gyr of the universe.
The lower limits on the lifetimes of
particles of dark matter are somewhat shorter,
for instance about \( 10^{18} \)  years
for 100 GeV particles~\cite{Bertone:2010zza},
and vary with their putative 
kinds~\cite{PhysRevD.93.103009, *PhysRevD.93.103517},
but they all tend to be much longer than \( 13.8 \) Gyr\@.
In what follows, I will assume for simplicity
that the matter in the universe is stable
on timescales that far exceed the age of the universe.
In that case, the mass of matter within a volume
\( a^3(t) \) remains constant, and so
the matter density varies with the 
scale factor \( a(t) \) as
\begin{equation}
\rho_m(t) = \rho_{m 0} \, \frac{a^3_0}{a^3(t)} .
\label {rhom at time t}
\end{equation}
\par
In terms of these densities,
the normalized Friedmann equation
(\ref {2d F eq}) is
\begin{equation}
   \begin{split}
\frac{H^2}{H^2_0} = {}&
\frac{\rho_\Lambda}{\rho_{c 0}} 
+ \frac{\rho_r}{\rho_{c 0}}
+ \frac{\rho_m}{\rho_{c 0}}
- \frac{k c^2}{a^2 H^2_0} \\
= {}&
\frac{\rho_{\Lambda 0}}{\rho_{c 0}} 
+ \frac{\rho_{r 0}}{\rho_{c 0}} \, \frac{a^4_0}{a^4}
+ \frac{\rho_{m 0}}{\rho_{c 0}}  \, \frac{a^3_0}{a^3}
- \frac{k c^2}{a^2_0 H^2_0} \, \frac{a^2_0}{a^2}.
\label {3d F eq}
   \end{split}
\end{equation}
These density ratios are called respectively
\( \Omega_{\Lambda 0} \),
\( \Omega_{r 0} \), 
\( \Omega_{m 0} \), and 
\( \Omega_{k 0} \equiv - k c^2/a^2_0 H^2_0 \)\@.
In terms of them
this formula (\ref {3d F eq}) for \( H^2/H^2_0 \) is
\begin{equation}
\frac{H^2}{H^2_0} = {}
\Omega_{\Lambda 0}
+ \Omega_{k 0}  \, \frac{a^2_0}{a^2}
+  \Omega_{m 0} \, \frac{a^3_0}{a^3} 
+  \Omega_{r 0} \, \frac{a^4_0}{a^4} .
\label {H^2/H^20}
\end{equation}
Since \( H = \dot a/a \), the element \( dt \) of time is
\( dt = H^{-1}_0 (da/a) (H_0/H) \),
and so setting \( x = a/a_0 \), we have
\begin{equation}
dt = \frac{1}{H_0} \, \frac{dx}{x} 
\, \frac{1}{\sqrt{\Omega_{\Lambda 0}
+ \Omega_{k 0} \, x^{-2}  
+ \Omega_{m 0} \, x^{-3}  
+ \Omega_{r 0} \, x^{-4} } } .
\label {dt=}
\end{equation}
Integrating, we get a 
formula~\cite{Weinberg2010.1.5}
for the time \( t(a/a_0) \) as a function
of the ratio \( a/a_0 \) of scale factors
\begin{equation}
t(a/a_0) =  \frac{1}{H_0} \int_0^{a/a_0}
\!\! \frac{dx}{\sqrt{ \Omega_{\Lambda 0} \, x^2
+ \Omega_{k 0} 
+ \Omega_{m 0} \, x^{-1}  
+ \Omega_{r 0} \, x^{-2} } }
\label {t(z)}
\end{equation}
in which the origin of time is at
at scale factor zero, \( t(0) = 0 \),
or equivalently at infinite redshift, \( z = (a_0/a) - 1 \)\@.
Light emitted at time \( t(a/a_0) = t(1/(z+1)) \)
reaches us with redshift \( z \)\@.

\begin{figure}[h]
\begin{center}
\includegraphics[trim=0 150 0 150, clip, width=\textwidth]{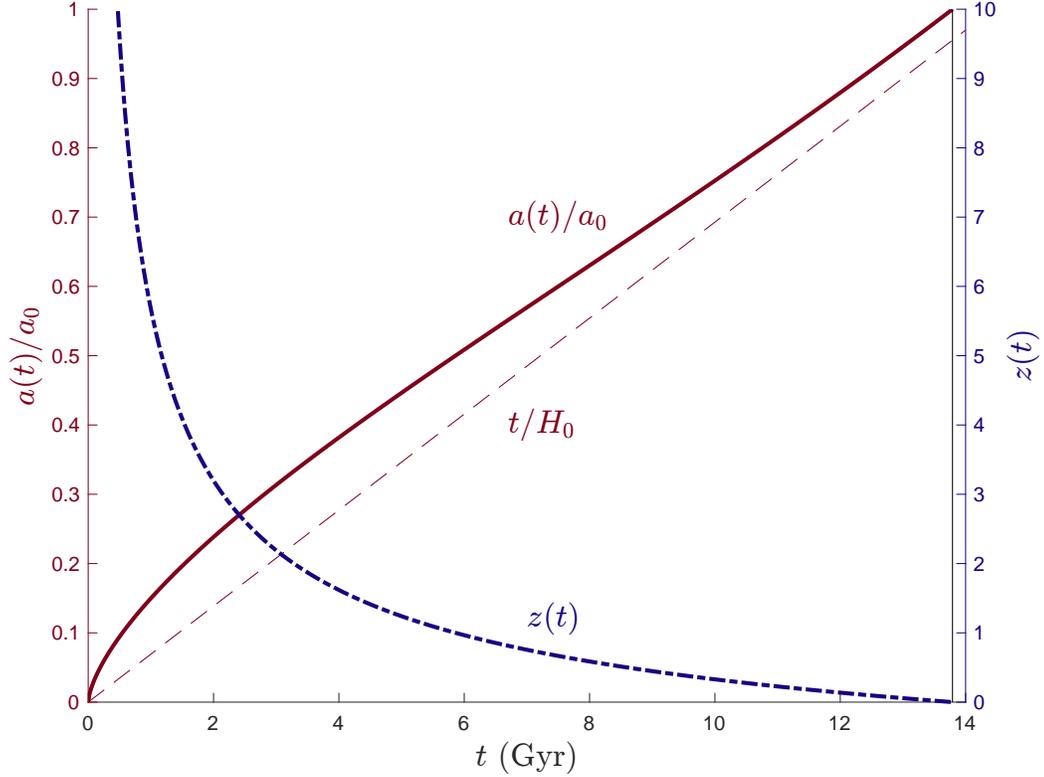}
\caption{The scale-factor ratio \( a(t)/a_0 \) 
(solid, red), 
the time \(t/H_0\)
(\ref {t(z)}, dashed, red),
and the redshift \( z(t) \) 
(dashdot, blue) are plotted against
of the time (\ref {t(z)}) in Gyr.
The vertical black line marks the present time.
A photon emitted with 
wavelength \( \lambda \)
at time \( t \) now has wavelength 
\( \lambda_0 = (a_0/a(t)) \, \lambda \)\@.}
\label{t2x}
\end{center}
\end{figure}

\par
I performed the integral (\ref {t(z)}) numerically
and plotted the scale-factor ratio \( a(t)/a_0 \) 
(solid, red)
and the redshift \( z(t) \)  
(dashdot, blue) 
in Figure~\ref{t2x} against
the time \( t \) in Gyr.
The ratio \( a(t)/a_0 \) is roughly 
the straight line \( t/H_0 \)
(dashed, red)\@.
The vertical black line marks
the present time.
The Fortran and Matlab codes for the 
figures of this paper are available
at arxiv.org as ancillary files.

\section{The end of the era of radiation
\label{The end of the era of radiation}}

\begin{figure}[h]
\begin{center}
\includegraphics[trim=0 150 0 150, clip, width=\textwidth]{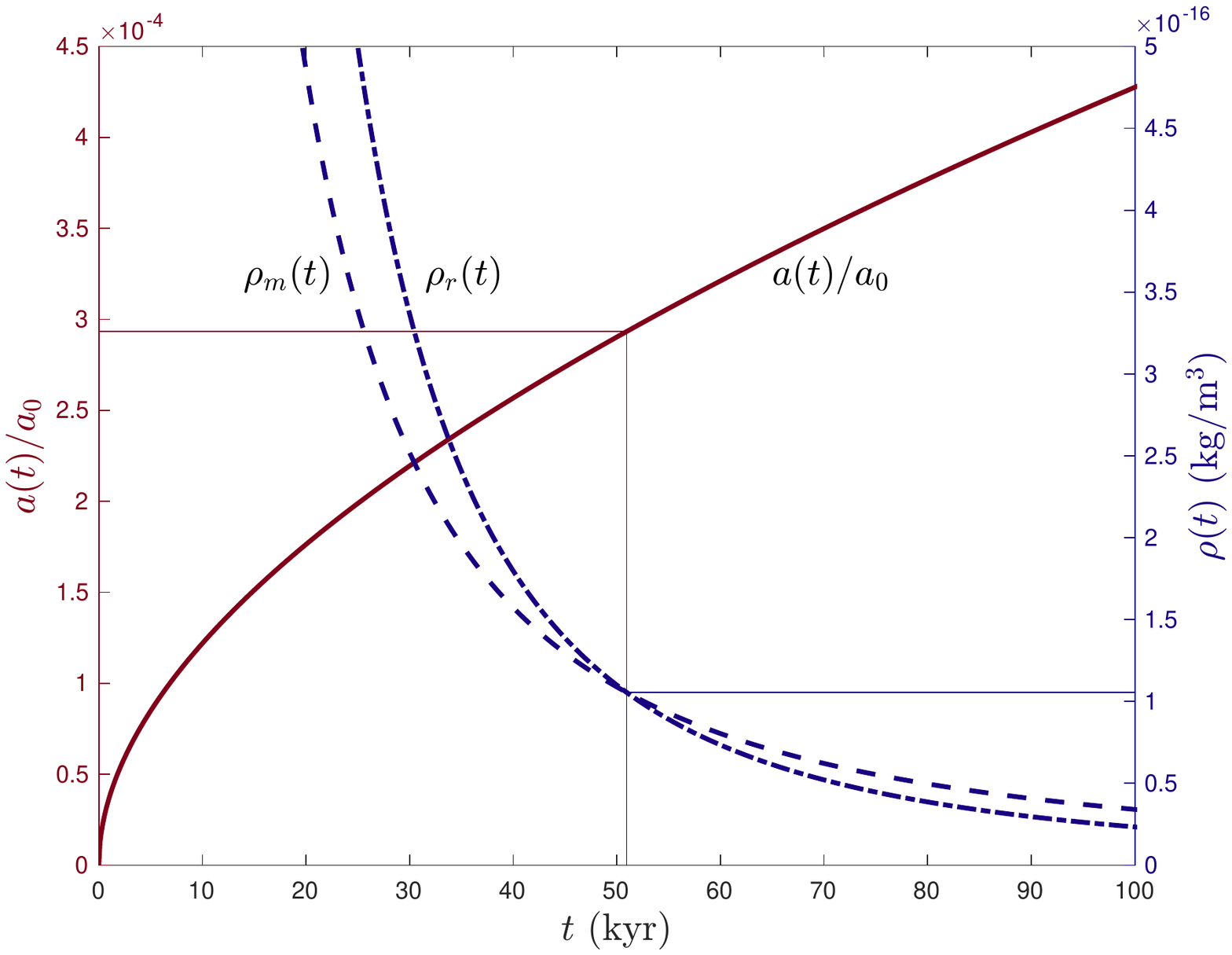}
\caption{The scale factor 
ratio \( a(t)/a_0 \) (solid, red),
the radiation density \( \rho_r \) (dashdot, blue),
and the matter density \( \rho_m \) (dashed, blue)
are plotted as functions 
of the time (\ref {t(z)}) in kyr.
The era of radiation ended
at \( t = 50,953 \) years 
(marked by the vertical line)
when the two densities were
equal to \( 1.055 \by 10^{-16} \) kg/m\({}^3 \)
(right horizontal line),
and \( a(t)/a_0 \) was \( 2.933 \by 10^{-4} \)
(left horizontal line)\@.}
\label {radiation}
\end{center}
\end{figure}

The era of radiation ended,
and the era of matter began 
when the density of radiation \( \rho_r(t) \),
decreasing rapidly (\ref {rhor at time t}) as \( 1/a^4(t) \),
dropped to that of matter \( \rho_m(t) \),
which fell more slowly (\ref {rhom at time t}) as \( 1/a^3(t) \)\@.
Their densities matched when
\begin{equation}
\rho_r(t) = \frac{a_0^4}{a^4(t)} \, \rho_{r,0} = \rho_m(t) = 
\frac{a_0^3}{a^3(t)} \, \rho_{m,0} 
\label{era of radiation ended}
\end{equation}
or when
\begin{equation}
\frac{a(t)}{a_0} = \frac{\rho_{r,0}}{\rho_{m,0}}
= \frac{\Omega_{r 0}}{\Omega_{m 0}}
= \frac{9.0606 \by 10^{-5}}{0.3089} 
= \left( 2.9332 \pm 0.0711 \right) \by 10^{-4} .
\label{a/a0 end of radiation}
\end{equation}
The corresponding redshift is
\begin{equation}
z = \frac{a_0}{a(t)} -1 = 3408.27 \pm  82.67 .
\label{redshift end of radiation}
\end{equation}
Using the scale-factor 
ratio (\ref{a/a0 end of radiation})
as the upper limit
of the integral (\ref {t(z)}), 
we find for the time at which
the end of the era of radiation 
ended 
\begin{equation}
t(2.9332 \by 10^{-4}) 
= 50,953 \pm 2236 \mbox{ years}
\end{equation}
after the time 
of infinite redshift.
In this estimate,
the standard deviation \( \sigma_t = 2236 \) years
reflects only the uncertainty in \( a(t)/a_0 \) 
and not the correlated uncertainties
in the other quantities 
in the integral (\ref {t(z)})\@.
\par
Figure~\ref {radiation} plots the scale-factor ratio
\( a(t)/a_0 \) against the time in kyr.
The vertical line in the figure
marks the time (50,953 years) 
of the transition from an era dominated
by radiation to one dominated by matter.
The left and right horizontal lines
mark the scale-factor ratio \( a(t)/a_0 = 2.933 \by 10^{-4} \)
and the equal densities \( 1.055 \by 10^{-16} \)
kg/m\({}^3\)\@.

\section{The end of the era of matter
\label{The end of the era of matter}}

\begin{figure}[h]
\begin{center}
\includegraphics[trim=0 150 0 150, clip, width=\textwidth]{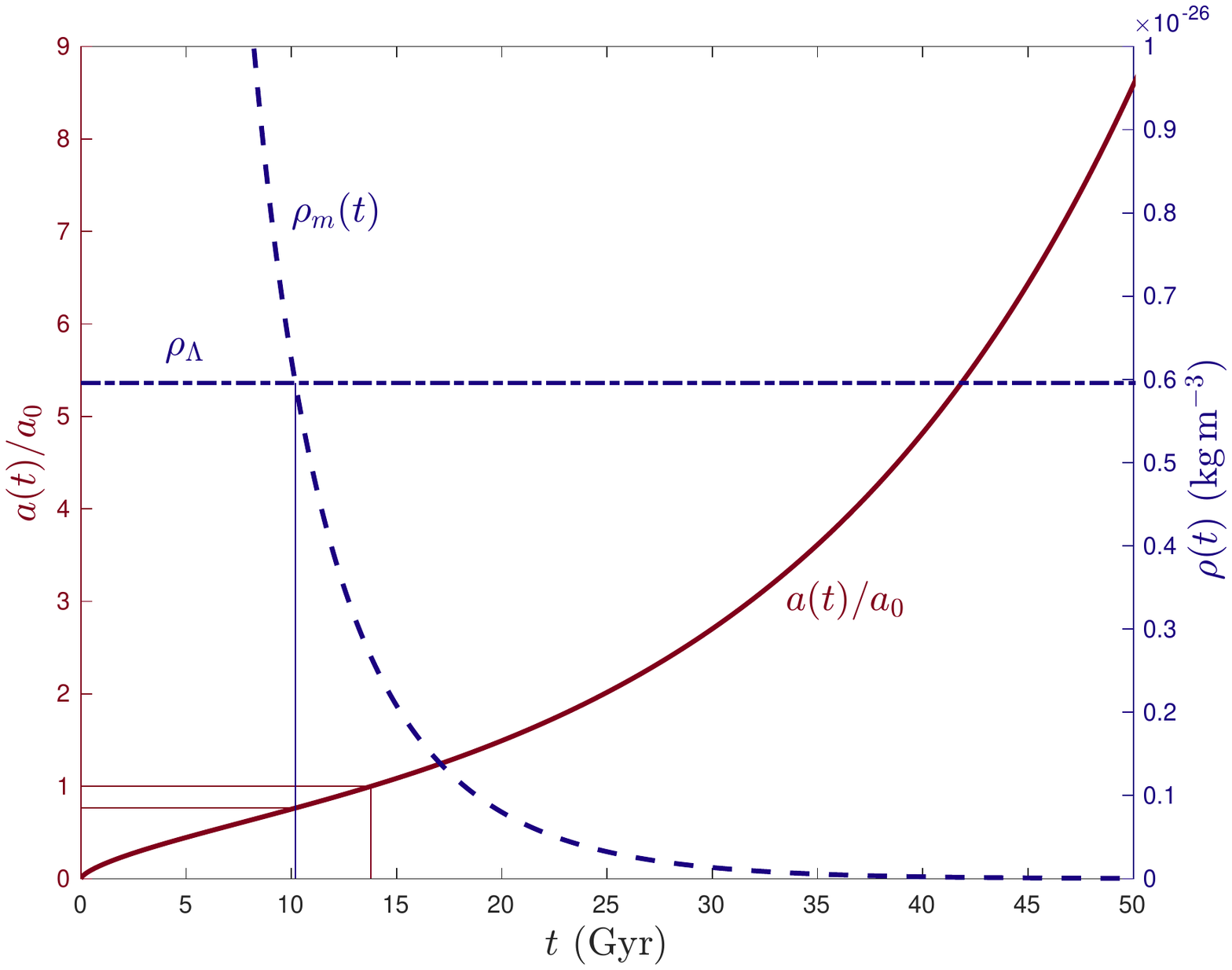}
\caption{The scale-factor ratio \( a(t)/a_0 \) (solid red),
the vacuum density \( \rho_\Lambda \) (dashdot, blue),
and the matter density \( \rho_m \) (dashed, blue)
are plotted as functions 
of the time (\ref {t(z)}) in Gyr 
since the Big Bang.
The era of matter ends
at \( t = 10.193 \) Gyr
(long vertical line)
when the two densities are equal.
The short vertical line marks
the present time 13.8 Gyr 
after which the falling matter density \( \rho_m(t) \)
and increasing significance of the vacuum density 
\( \rho_{\Lambda 0} \) cause the 
scale-factor ratio \( a(t)/a_0 \) to increase
exponentially.}
\label {vacuum}
\end{center}
\end{figure}

\par
The era of matter ended,
and the era of dark energy began 
when the mass density of matter \( \rho_m(t) \),
decreasing (\ref {rhom at time t}) as \( 1/a^3(t) \),
dropped to the constant 
mass density \( \rho_\Lambda \)
of the vacuum (\ref {rhov})\@.
Their densities matched when
\( \rho_m(t) = \rho_\Lambda =  \rho_{\Lambda 0}\)
which occurred when 
\begin{equation}
\rho_m(t) = \frac{a_0^3}{a^3(t)} \, \rho_{m,0}  = \rho_{\Lambda 0}
\label{end of matter era}
\end{equation}
or when
\begin{equation}
\frac{a}{a_0} = \lt( \frac{\rho_{m 0}}{\rho_{\Lambda 0}} \rt)^{1/3}
= \lt( \frac{\Omega_{m 0}}{\Omega_{\Lambda 0}} \rt)^{1/3}
= \lt( \frac{0.3089}{0.6911} \rt)^{1/3} 
= 0.76458 \pm .01681 . 
\label{a/a_0}
\end{equation}
The corresponding redshift is
\begin{equation}
z =\frac{a_0}{a} - 1 = 0.3079 \pm .0288 . 
\label{z at end of matter era}
\end{equation}
The integral (\ref {t(z)})
gives the time \( t(0.3079) \) as 
\begin{equation}
   \begin{split}
t(0.76458) = {}& \frac{1}{H_0} \int_0^{0.76458}
\!\! \frac{dx}{ \sqrt{\Omega_{\Lambda 0} \, x^2
+ \Omega_{k 0} 
+ \Omega_{m 0} \, x^{-1}  
+ \Omega_{r 0} \, x^{-2} } } \\
={}&  \left( 10.1928 \pm 0.0375 \right)\mbox{ Gyr} 
\label {t(0.3079)}
   \end{split}
\end{equation}
in which the quoted uncertainty 
reflects only that of \( a(t)/a_0 \)\@.
The era of dark energy began 
about \( 13.799 - 10.193 = 3.606 \) billion years ago.
\par
The same integral (\ref {t(z)}) gives
the age of the universe as 
its value at \( a(t) = a_0 \) which is
\( t(1) ={} 13.792 \) Gyr;
the Planck value is
\( ( 13.799 \pm 0.021 ) \) Gyr. 
\par
Figure~\ref {vacuum} plots the 
scale-factor ratio \( a(t)/a_0 \)
(solid, red), the matter density
\( \rho_m(t) \) (dashed, blue), and
the vacuum density \( \rho_\Lambda \)
(dashdot, blue) against the time in Gyr.
The long vertical line marks the present time
13.8 Gyr.  
The exponential growth of 
the scale-factor ratio \( a(t)/a_0 \)
due to the falling density ratio \( \rho_m(t) / \rho_\Lambda \)
becomes evident after 30 Gyr.

\section {The luminosity distance
\label  {The luminosity distance}}

The Robinson-Walker invariant line element 
of an isotropic and homogeneous universe is
\begin{equation}
ds^2 ={} a^2(t) \left[ \frac{dr^2}{1 - k r^2} 
+ r^2 \left( d\theta^2 + \sin^2\theta d\phi^2 \right)
\right] - c^2 dt^2 .
\label {The Robinson-Walker invariant line element} 
\end{equation}
A photon 
leaves a source at comoving
coordinate \( r \) at time t and comes
radially (\( d\theta = d\phi = 0 \))
along a path with \( ds^2 = 0 \)
to us at \( r = 0 \) 
and time \( t_0 \)\@.
Since  \( c^2 dt^2 = a^2(t) dr^2 / (1 - k r^2 ) \),
the time \( t < t_0 \) and
radial coordinate \( r > 0 \) 
of emission satisfy
\begin{equation} 
\int_t^{t_0} \frac{c \, dt'}{a(t')} =
\int_0^r \frac{dr'}{\sqrt{1 - k r'^2}} 
= \left\{
   \begin{array} {lr} 
       \arcsin r & \quad k = 1 \\
       r & \quad k = 0 \\
        \arcsinh r & \quad k = -1 
   \end{array}
\right. .
\label {the time t and r obey}
\end{equation}
As in our derivation
of the formula (\ref{dt=}) for \( dt \),
we set \( x = a(t)/a_0 \) and use
\( dt / a(t) = da/(a^2(t) H(t)) = dx/(a_0 x^2 H(t) ) \)
to get~\cite{Weinberg2010.1.5}
\begin{equation}
\left. 
\begin{array} {lr} 
 k = 1  &     \arcsin r  \\
 k = 0  &     r   \\
 k = -1 &    \arcsinh r
   \end{array}
   \right\} ={}
    \frac{c}{a_0 H_0} \int^1_{a/a_0}
\frac{dx}{x^2 \sqrt{\Omega_{\Lambda 0}
+ \Omega_{k 0} \, x^{-2}  
+ \Omega_{m 0} \, x^{-3}  
+ \Omega_{r 0} \, x^{-4} } }  .
\label {integral for r}   
\end{equation}

\par
If \( L \) is the absolute luminosity
(or power) of a source,
then at a short distance \( r \)
its apparent luminosity 
is \( L_a = L/(4 \pi r^2)  \)\@.
At longer distances, one must
expand \( r \) to \( a_0 \, r \)
and also include two factors
of \( a(t)/a_0 \), one
to adjust the emission and arrival rates
of the photons, and the other to
adjust the emitted and arriving energies
of the photons~\cite{Weinberg2010.1.4}\@.
For a source at any comoving coordinate \( r \),
the apparent luminosity is
\begin{equation}
L_a ={} \frac{L \, a^2(t)}{4 \pi \, r^2 \, a^4_0} 
= \frac{L }{4 \pi \, r^2 \, a_0^2 \, (z + 1)^2 }  
\equiv \frac{L }{4 \pi \, d^2_L }  
\label {apparent luminosity}
\end{equation}
in which \( d_L \)
is the luminosity distance
\begin{equation}
d_L(z) = r(z) \, a_0 \, ( 1 + z ) .
\label {luminosity distance}
\end{equation}
\par
For a flat (\( k = 0 \)) universe,
the comoving coordinate \( r \)
is the middle equation of the
triplet (\ref {integral for r}),
and so with \( \Omega_k = 0 \) 
the luminosity distance is
\begin{equation}
d_L^{(0)}(z) ={} \frac{c \, (1+z)}{H_0} \int^1_{1/(1+z)} 
\frac{dx}{\sqrt{\Omega_{\Lambda 0} \, x^4
+ \Omega_{m 0} \, x  
+ \Omega_{r 0} } }  .
\label {the k = 0 luminosity distance}
\end{equation}
For a closed (\( k = 1 \)) universe,
the comoving coordinate \( r \)
is the top equation of the
triplet (\ref {integral for r}),
and so with \( c/(a_0 H_0) = \sqrt{|\Omega_k|} \)
the luminosity distance is
\begin{equation}
d_L^{(1)}(z) ={} \frac{ c \, (1+z) }{H_0  \sqrt{|\Omega_k|}}
\sin \left( \sqrt{|\Omega_k|}
\int^1_{1/(1+z)} 
\frac{dx}{\sqrt{\Omega_{\Lambda 0} \, x^4
+ \Omega_{k 0} \, x^2 
+ \Omega_{m 0} \, x 
+ \Omega_{r 0} } }  
\right) .
\label {the k = 1 luminosity distance}
\end{equation}
For an open (\( k = - 1 \)) universe,
the comoving coordinate \( r \)
is the bottom equation of the
triplet (\ref {integral for r}),
and so with \( c/(a_0 H_0) = \sqrt{\Omega_k} \)
the luminosity distance is
\begin{equation}
d_L^{(-1)}(z) ={} \frac{ c \, (1+z) }{H_0  \sqrt{|\Omega_k|}}
\sinh \left( \sqrt{\Omega_k}
\int^1_{1/(1+z)}
\frac{dx}{\sqrt{\Omega_{\Lambda 0} \, x^4
+ \Omega_{k 0} \, x^2 
+ \Omega_{m 0} \, x
+ \Omega_{r 0} } }  
\right) .
\label {the k = - 1 luminosity distance}
\end{equation}

\begin{figure}[h]
\begin{center}
\includegraphics[trim=0 150 0 150, clip, width=\textwidth]{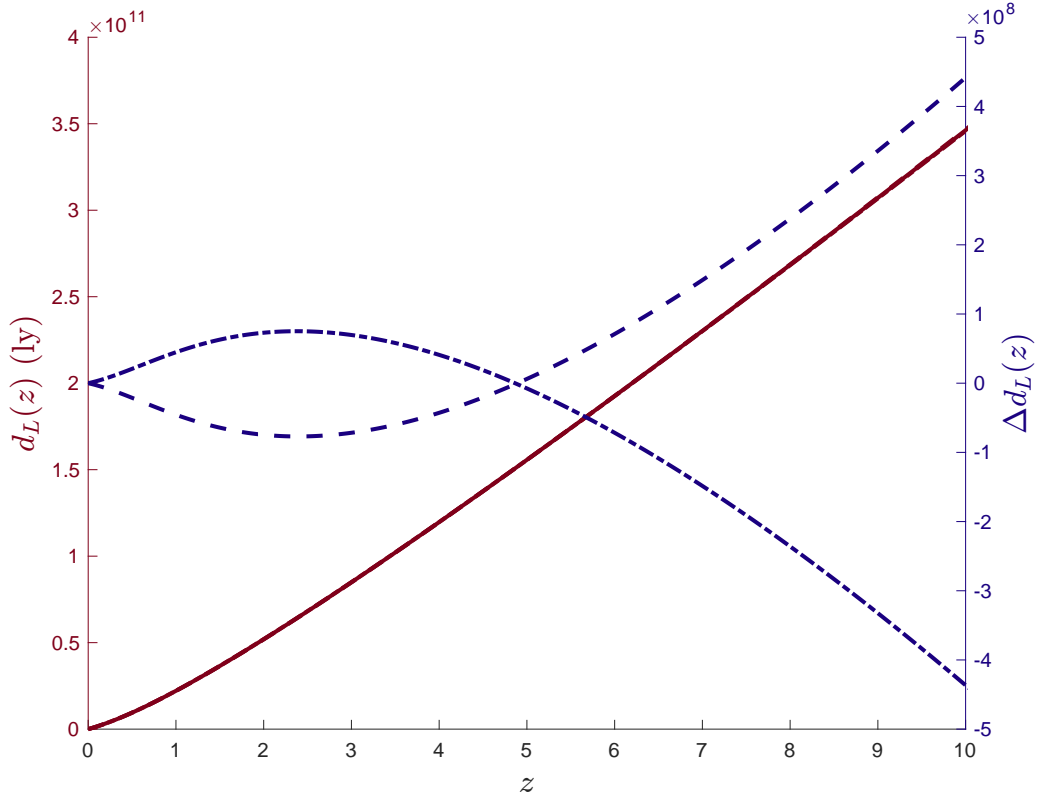}
\caption{The luminosity distance \( d^{(-1)}_L(z) \)
(\ref {the k = - 1 luminosity distance})
as given by the Planck data of Table~\ref {table}
with \( \Omega_k = 0.0008 \)
is plotted (solid, red) against the redshift \( z \)
for \( 0 < z < 10 \)\@.
Also plotted are the 
most open (\( d^{(-1)}_L(z, \Omega_k = 0.0048) \),
\ref  {the k = - 1 luminosity distance}, red dashed)
and most closed 
(\( d^{(1)}_L(z, \Omega_k = - 0.0031) \),
\ref  {the k = 1 luminosity distance}, red dashdot)
luminosity distances allowed by the Planck data
up to one standard deviation (68\%)\@.
Because the three curves nearly overlap, 
their differences 
(\ref {d - dopen}, blue dashed)
and 
(\ref {d - dclosed}, blue dashdot)
also are plotted.}
\label {short distances}
\end{center}
\end{figure}

\begin{figure}[h]
\begin{center}
\includegraphics[trim=0 150 0 150, clip, width=\textwidth]{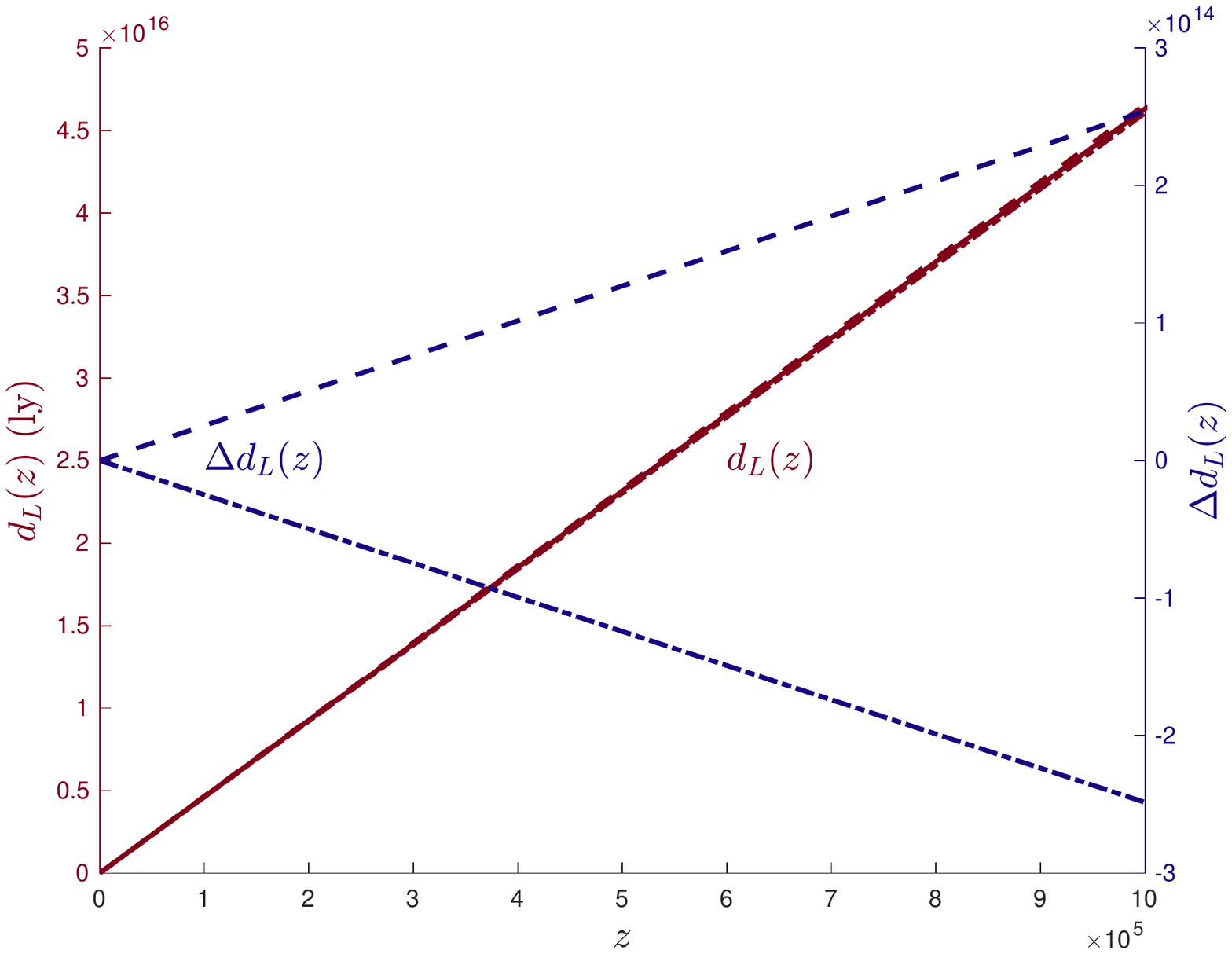}
\caption{The luminosity distance \( d^{(-1)}_L(z) \)
(\ref {the k = - 1 luminosity distance})
as given by the Planck data of Table~\ref {table}
with \( \Omega_k = 0.0008 \)
is plotted (solid, red) against the redshift \( z \)
for \( 0 < z < 10^6 \)\@.
Also plotted are the 
most open (\( d^{(-1)}_L(z, \Omega_k = 0.0048) \),
\ref  {the k = - 1 luminosity distance}, red dashed)
and most closed 
(\( d^{(1)}_L(z, \Omega_k = - 0.0031) \),
\ref  {the k = 1 luminosity distance}, red dashdot)
luminosity distances
allowed to 68\%\@.
The three curves closely follow
the big-\(z\) formula (\ref {d_L for big z}), and
because they nearly overlap, 
their differences 
(\ref {d - dopen}, blue dashed)
and 
(\ref {d - dclosed}, blue dashdot)
also are plotted.}
\label {distances}
\end{center}
\end{figure}

\par
The luminosity distance  \( d^{(-1)}_L(z) \) 
(\ref {the k = - 1 luminosity distance})
as given by the Planck data of Table~\ref {table}
and by the value 
\( \Omega_{r 0} = 0.90606 \by 10^{-4} \) 
of equation (\ref {Omegar})
is plotted 
(solid, red) against the redshift \( z \)
for moderate \( z < 10 \) in figure~\ref {short distances}
and for \( z < 10^6 \) in figure~\ref {distances}\@.
Also plotted in both figures are the 
most-open luminosity distance
(\( d^{(-1)}_L(z, \Omega_k = 0.0048) \),
\ref  {the k = - 1 luminosity distance}, red dashed)
and the most-closed luminosity distance
(\( d^{(1)}_L(z, \Omega_k = - 0.0031) \),
\ref  {the k = 1 luminosity distance}, red dashdot)
allowed to 68\% by the Planck data.
Because the limits 
(\( | \Omega_k - 0.0008 | < 0.004 \))
on \( \Omega_k \) are so tight,
all three luminosity distances 
are nearly indistinguishable in the figures.
So the figures also plot the difference between the
most-open luminosity distance 
and the most likely luminosity distance 
(blue, dashed)
\begin{equation}
\Delta d_L(z) = {}
d^{(-1)}_L(z, \Omega_k = 0.0048) - d^{(-1)}_L(z)
\label {d - dopen}
\end{equation}
as well as the difference between the
most-closed luminosity distance 
and the most likely luminosity distance
(blue, dashdot)
\begin{equation}
\Delta d_L(z) = {}
d^{(1)}_L(z, \Omega_k = - 0.0031) - d^{(-1)}_L(z) .
\label {d - dclosed}
\end{equation}
These relative differences 
\( | \Delta d_L(z) / d_L(z) | < 6 \by 10^{-3} \) 
are less than 1\%\@.
\par
The argument of the sinh
in the formula (\ref {the k = - 1 luminosity distance})
for the luminosity distance as given by
the Planck values of the density ratios
approaches 0.09063 as \( z \to \infty \), 
and so for large redshifts
the luminosity distance \( d_L^{(-1)} (z) \) rises as 
\begin{equation}
d_L^{(-1)} (z) ={} 3.2087  \, \frac{ c \, (1+z) }{H_0} 
\label {d_L for big z}
\end{equation}
as verified by figure~\ref {distances}\@.

\begin{acknowledgments}
I must thank Shashank Shalgar
for many helpful conversations.
This work also was advanced
by discussions with and e-mail from
Franco Giuliani and Rouzbeh Allahverdi.
I should also like to thank
Jooyoung Lee for inviting
me to the Korea Institute
for Advanced Study (KIAS)
where some of this work was done.
\end{acknowledgments}
\bibliography{physics}

\begin{thebibliography}{15}%
\makeatletter
\providecommand \@ifxundefined [1]{%
 \@ifx{#1\undefined}
}%
\providecommand \@ifnum [1]{%
 \ifnum #1\expandafter \@firstoftwo
 \else \expandafter \@secondoftwo
 \fi
}%
\providecommand \@ifx [1]{%
 \ifx #1\expandafter \@firstoftwo
 \else \expandafter \@secondoftwo
 \fi
}%
\providecommand \natexlab [1]{#1}%
\providecommand \enquote  [1]{``#1''}%
\providecommand \bibnamefont  [1]{#1}%
\providecommand \bibfnamefont [1]{#1}%
\providecommand \citenamefont [1]{#1}%
\providecommand \href@noop [0]{\@secondoftwo}%
\providecommand \href [0]{\begingroup \@sanitize@url \@href}%
\providecommand \@href[1]{\@@startlink{#1}\@@href}%
\providecommand \@@href[1]{\endgroup#1\@@endlink}%
\providecommand \@sanitize@url [0]{\catcode `\\12\catcode `\$12\catcode
  `\&12\catcode `\#12\catcode `\^12\catcode `\_12\catcode `\%12\relax}%
\providecommand \@@startlink[1]{}%
\providecommand \@@endlink[0]{}%
\providecommand \url  [0]{\begingroup\@sanitize@url \@url }%
\providecommand \@url [1]{\endgroup\@href {#1}{\urlprefix }}%
\providecommand \urlprefix  [0]{URL }%
\providecommand \Eprint [0]{\href }%
\providecommand \doibase [0]{http://dx.doi.org/}%
\providecommand \selectlanguage [0]{\@gobble}%
\providecommand \bibinfo  [0]{\@secondoftwo}%
\providecommand \bibfield  [0]{\@secondoftwo}%
\providecommand \translation [1]{[#1]}%
\providecommand \BibitemOpen [0]{}%
\providecommand \bibitemStop [0]{}%
\providecommand \bibitemNoStop [0]{.\EOS\space}%
\providecommand \EOS [0]{\spacefactor3000\relax}%
\providecommand \BibitemShut  [1]{\csname bibitem#1\endcsname}%
\let\auto@bib@innerbib\@empty
\bibitem [{\citenamefont {Ade}\ \emph {et~al.}(2015)\citenamefont {Ade} \emph
  {et~al.}}]{Ade:2015xua}%
  \BibitemOpen
  \bibfield  {author} {\bibinfo {author} {\bibfnamefont {P.~A.~R.}\
  \bibnamefont {Ade}} \emph {et~al.} (\bibinfo {collaboration} {Planck}),\
  }\href@noop {} {\  (\bibinfo {year} {2015})},\ \Eprint
  {http://arxiv.org/abs/1502.01589} {arXiv:1502.01589 [astro-ph.CO]}
  \BibitemShut {NoStop}%
\bibitem [{\citenamefont {{Fixsen}}(2009)}]{2009ApJ...707..916F}%
  \BibitemOpen
  \bibfield  {author} {\bibinfo {author} {\bibfnamefont {D.~J.}\ \bibnamefont
  {{Fixsen}}},\ }\href {\doibase 10.1088/0004-637X/707/2/916} {\bibfield
  {journal} {\bibinfo  {journal} {\apj}\ }\textbf {\bibinfo {volume} {707}},\
  \bibinfo {pages} {916} (\bibinfo {year} {2009})},\ \Eprint
  {http://arxiv.org/abs/0911.1955} {arXiv:0911.1955} \BibitemShut {NoStop}%
\bibitem [{\citenamefont {Cahill}(2013{\natexlab{a}})}]{Cahill2013.4.10}%
  \BibitemOpen
  \bibfield  {author} {\bibinfo {author} {\bibfnamefont {K.}~\bibnamefont
  {Cahill}},\ }\enquote {\bibinfo {title} {\textsl{Physical Mathematics}},}\ \
  (\bibinfo  {publisher} {Cambridge University Press},\ \bibinfo {year}
  {2013})\ Chap.\ \bibinfo {chapter} {4.10}, pp.\ \bibinfo {pages}
  {149--151}\BibitemShut {NoStop}%
\bibitem [{\citenamefont {Weinberg}(2010{\natexlab{a}})}]{Weinberg2010.2.1}%
  \BibitemOpen
  \bibfield  {author} {\bibinfo {author} {\bibfnamefont {S.}~\bibnamefont
  {Weinberg}},\ }\enquote {\bibinfo {title} {\textsl{Cosmology}},}\ \ (\bibinfo
   {publisher} {Oxford University Press},\ \bibinfo {year} {2010})\ Chap.\
  \bibinfo {chapter} {2.1}, pp.\ \bibinfo {pages} {105--107}\BibitemShut
  {NoStop}%
\bibitem [{\citenamefont {Cahill}(2013{\natexlab{b}})}]{Cahill2013.11.48}%
  \BibitemOpen
  \bibfield  {author} {\bibinfo {author} {\bibfnamefont {K.}~\bibnamefont
  {Cahill}},\ }\enquote {\bibinfo {title} {\textsl{Physical Mathematics}},}\ \
  (\bibinfo  {publisher} {Cambridge University Press},\ \bibinfo {year}
  {2013})\ Chap.\ \bibinfo {chapter} {11.48}, pp.\ \bibinfo {pages}
  {457--463}\BibitemShut {NoStop}%
\bibitem [{\citenamefont {Weinberg}(2010{\natexlab{b}})}]{Weinberg2010.1.5}%
  \BibitemOpen
  \bibfield  {author} {\bibinfo {author} {\bibfnamefont {S.}~\bibnamefont
  {Weinberg}},\ }\enquote {\bibinfo {title} {\textsl{Cosmology}},}\ \ (\bibinfo
   {publisher} {Oxford University Press},\ \bibinfo {year} {2010})\ Chap.\
  \bibinfo {chapter} {1.5}, pp.\ \bibinfo {pages} {34--45}\BibitemShut
  {NoStop}%
\bibitem [{\citenamefont {Weinberg}(2010{\natexlab{c}})}]{Weinberg2010.1.12}%
  \BibitemOpen
  \bibfield  {author} {\bibinfo {author} {\bibfnamefont {S.}~\bibnamefont
  {Weinberg}},\ }\enquote {\bibinfo {title} {\textsl{Cosmology}},}\ \ (\bibinfo
   {publisher} {Oxford University Press},\ \bibinfo {year} {2010})\ Chap.\
  \bibinfo {chapter} {1.12}, pp.\ \bibinfo {pages} {89--98}\BibitemShut
  {NoStop}%
\bibitem [{\citenamefont {Peebles}\ and\ \citenamefont
  {Ratra}(2003)}]{Peebles:2002gy}%
  \BibitemOpen
  \bibfield  {author} {\bibinfo {author} {\bibfnamefont {P.~J.~E.}\
  \bibnamefont {Peebles}}\ and\ \bibinfo {author} {\bibfnamefont
  {B.}~\bibnamefont {Ratra}},\ }\href {\doibase 10.1103/RevModPhys.75.559}
  {\bibfield  {journal} {\bibinfo  {journal} {Rev. Mod. Phys.}\ }\textbf
  {\bibinfo {volume} {75}},\ \bibinfo {pages} {559} (\bibinfo {year} {2003})},\
  \Eprint {http://arxiv.org/abs/astro-ph/0207347} {arXiv:astro-ph/0207347
  [astro-ph]} \BibitemShut {NoStop}%
\bibitem [{\citenamefont {Linder}(2008)}]{Linder:2007wa}%
  \BibitemOpen
  \bibfield  {author} {\bibinfo {author} {\bibfnamefont {E.~V.}\ \bibnamefont
  {Linder}},\ }\href {\doibase 10.1007/s10714-007-0550-z} {\bibfield  {journal}
  {\bibinfo  {journal} {Gen. Rel. Grav.}\ }\textbf {\bibinfo {volume} {40}},\
  \bibinfo {pages} {329} (\bibinfo {year} {2008})},\ \Eprint
  {http://arxiv.org/abs/0704.2064} {arXiv:0704.2064 [astro-ph]} \BibitemShut
  {NoStop}%
\bibitem [{\citenamefont {Araki}\ \emph {et~al.}(2006)\citenamefont {Araki}
  \emph {et~al.}}]{Araki:2005jt}%
  \BibitemOpen
  \bibfield  {author} {\bibinfo {author} {\bibfnamefont {T.}~\bibnamefont
  {Araki}} \emph {et~al.} (\bibinfo {collaboration} {KamLAND}),\ }\href
  {\doibase 10.1103/PhysRevLett.96.101802} {\bibfield  {journal} {\bibinfo
  {journal} {Phys. Rev. Lett.}\ }\textbf {\bibinfo {volume} {96}},\ \bibinfo
  {pages} {101802} (\bibinfo {year} {2006})},\ \Eprint
  {http://arxiv.org/abs/hep-ex/0512059} {arXiv:hep-ex/0512059 [hep-ex]}
  \BibitemShut {NoStop}%
\bibitem [{\citenamefont {Ahmed}\ \emph {et~al.}(2004)\citenamefont {Ahmed}
  \emph {et~al.}}]{PhysRevLett.92.102004}%
  \BibitemOpen
  \bibfield  {author} {\bibinfo {author} {\bibfnamefont {S.}~\bibnamefont
  {Ahmed}} \emph {et~al.} (\bibinfo {collaboration} {SNO Collaboration}),\
  }\href {\doibase 10.1103/PhysRevLett.92.102004} {\bibfield  {journal}
  {\bibinfo  {journal} {Phys. Rev. Lett.}\ }\textbf {\bibinfo {volume} {92}},\
  \bibinfo {pages} {102004} (\bibinfo {year} {2004})}\BibitemShut {NoStop}%
\bibitem [{\citenamefont {Silk}\ \emph {et~al.}(2010)\citenamefont {Silk} \emph
  {et~al.}}]{Bertone:2010zza}%
  \BibitemOpen
  \bibfield  {author} {\bibinfo {author} {\bibfnamefont {J.}~\bibnamefont
  {Silk}} \emph {et~al.},\ }\href
  {http://www.cambridge.org/uk/catalogue/catalogue.asp?isbn=9780521763684}
  {\emph {\bibinfo {title} {{Particle Dark Matter: Observations, Models and
  Searches}}}},\ edited by\ \bibinfo {editor} {\bibfnamefont {G.}~\bibnamefont
  {Bertone}}\ (\bibinfo  {publisher} {Cambridge University Press (2010)},\
  \bibinfo {year} {2010})\BibitemShut {NoStop}%
\bibitem [{\citenamefont {Baring}\ \emph {et~al.}(2016)\citenamefont {Baring},
  \citenamefont {Ghosh}, \citenamefont {Queiroz},\ and\ \citenamefont
  {Sinha}}]{PhysRevD.93.103009}%
  \BibitemOpen
  \bibfield  {author} {\bibinfo {author} {\bibfnamefont {M.~G.}\ \bibnamefont
  {Baring}}, \bibinfo {author} {\bibfnamefont {T.}~\bibnamefont {Ghosh}},
  \bibinfo {author} {\bibfnamefont {F.~S.}\ \bibnamefont {Queiroz}}, \ and\
  \bibinfo {author} {\bibfnamefont {K.}~\bibnamefont {Sinha}},\ }\href
  {\doibase 10.1103/PhysRevD.93.103009} {\bibfield  {journal} {\bibinfo
  {journal} {Phys. Rev. D}\ }\textbf {\bibinfo {volume} {93}},\ \bibinfo
  {pages} {103009} (\bibinfo {year} {2016})}\BibitemShut {NoStop}%
\bibitem [{\citenamefont {Lu}\ and\ \citenamefont
  {Zong}(2016)}]{PhysRevD.93.103517}%
  \BibitemOpen
  \bibfield  {author} {\bibinfo {author} {\bibfnamefont {B.-Q.}\ \bibnamefont
  {Lu}}\ and\ \bibinfo {author} {\bibfnamefont {H.-S.}\ \bibnamefont {Zong}},\
  }\href {\doibase 10.1103/PhysRevD.93.103517} {\bibfield  {journal} {\bibinfo
  {journal} {Phys. Rev. D}\ }\textbf {\bibinfo {volume} {93}},\ \bibinfo
  {pages} {103517} (\bibinfo {year} {2016})}\BibitemShut {NoStop}%
\bibitem [{\citenamefont {Weinberg}(2010{\natexlab{d}})}]{Weinberg2010.1.4}%
  \BibitemOpen
  \bibfield  {author} {\bibinfo {author} {\bibfnamefont {S.}~\bibnamefont
  {Weinberg}},\ }\enquote {\bibinfo {title} {\textsl{Cosmology}},}\ \ (\bibinfo
   {publisher} {Oxford University Press},\ \bibinfo {year} {2010})\ Chap.\
  \bibinfo {chapter} {1.4}, pp.\ \bibinfo {pages} {31--34}\BibitemShut
  {NoStop}%
\end{thebibliography}%
\end{document}